\documentclass[11pt]{article}
\usepackage{latexsym}  
\usepackage{amssymb}
\usepackage{graphicx}
\usepackage{amsmath}

\newcommand{\be}{\begin{equation}}
\newcommand{\ee}{\end{equation}}
\newcommand{\bea}{\begin{eqnarray}}
\newcommand{\eea}{\end{eqnarray}}
\newcommand{\bref}[1]{(\ref{#1})}
\newcommand{\la}{\langle}
\newcommand{\ra}{\rangle}
\begin{document}
\begin{titlepage}
\begin{flushright}
\today
\end{flushright}
\vspace{4\baselineskip}
\begin{center}
{\Large\bf  Model independent constraints of the averaged neutrino masses revisited.}
\end{center}
\vspace{1cm}
\begin{center}
{\large Takeshi Fukuyama$^{a,}$
\footnote{E-mail:fukuyama@se.ritsumei.ac.jp}}
and
{\large Hiroyuki Nishiura$^{b,}$
\footnote{E-mail:nishiura@is.oit.ac.jp}}
\end{center}
\vspace{0.2cm}
\begin{center}
${}^{a}$ {\small \it Research Center for Nuclear Physics (RCNP),
Osaka University, Ibaraki, Osaka, 567-0047, Japan}\\[.2cm]

${}^{b} $ {\small \it Faculty of Information Science and Technology, 
Osaka Institute of Technology,\\ Hirakata, Osaka 573-0196, Japan}

\vskip 10mm
\end{center}
\vskip 10mm
\begin{abstract}
Averaged neutrino masses defined by $\la m_\nu\ra_{ab} \equiv\left| \sum_{j=1}^{3}U_{aj}U_{bj}m_j\right|$ ($a,b=e,\mu,\tau$) 
are reanalyzed using up-to-date observed MNS parameters and neutrino masses 
by the neutrino oscillation experiments together with the cosmological constraint on neutrino masses.
The values of $\la m_\nu\ra_{ab}$ are model-independently evaluated in terms of effective neutrino mass defined by $\overline{m_\nu}\equiv\left(\sum |U_{ej}|^2m_j^2\right)^{1/2}$ which is observable in the single beta decay.  
We obtain lower bound for $\langle m_\nu \rangle_{ee}$ in the inverted hierarchy(IH) case,  
$17~\mbox{meV } \leq \langle m_\nu \rangle_{ee}$ and one for $\langle m_\nu \rangle_{\tau \mu}$ in the normal hierarchy(NH) case, $5~\mbox{meV }\leq \langle m_\nu \rangle_{\tau \mu}$. 
We also obtain that all the averaged masses $\la m_\nu\ra _{ab}$ have upper bounds which are at most $80~\mbox{meV}$. 
\end{abstract}
\end{titlepage}
The parameters of Maki-Nakagawa-Sakata (MNS) lepton mixing matrix \cite{MNS} \cite{Pontecorvo}
have been determined except for CP violating phases from the neutrino oscillation experiments. However, 
the absolute values of the neutrino masses $m_i$, the Dirac, and the Majorana CP violating phases are still remained undetermined.
Neutrino masses are also constrained from many other different experiments such as
$\beta$ decay, neutrinoless double $\beta$ decay experiments, and cosmological observations etc.

The $\beta$ decay restricts the effective neutrino mass defined by
\be
\overline{m_\nu}=\left(\sum |U_{ej}|^2m_j^2\right)^{1/2}
\label{mbeta}
\ee
and its experimental upper limits at 95\% C.L. are  
\bea
&&\overline{m_\nu}<2.3 ~\mbox{ eV}~~\mbox{(Mainz~\cite{Mainz})},\\
&&\overline{m_\nu}<2.1 ~\mbox{ eV}~~\mbox{(Troitsk~\cite{Troitsk})}.
\eea
Here the MNS matrix $U$ is represented in the standard form,
\begin{equation}
U=
\left(
\begin{array}{ccc}
c_{12}c_{13}&s_{12}c_{13}&s_{13}e^{-i\delta }\\
-s_{12}c_{23}-c_{12}s_{13}s_{23}e^{i\delta }&
c_{12}c_{23}-s_{12}s_{13}s_{23}e^{i\delta}&c_{13}s_{23}\\
s_{12}s_{23}-c_{12}s_{13}c_{23}e^{i\delta }&
-c_{12}s_{23}-s_{12}s_{13}c_{23}e^{i\delta }&c_{13}c_{23}\\
\end{array}
\right)
\left(
\begin{array}{ccc}
1 & 0 & 0\\
0 & e^{i\beta} & 0\\
0 & 0 & e^{i\gamma}\\
\end{array}
\right). \label{CKM}
\end{equation}
Here $c_{ij}=\cos\theta_{ij}$ and $s_{ij}=\sin\theta_{ij}$ with $\theta_{ij}$ being lepton mixing angles. 
The $\delta $ is the Dirac CP violating phase and the $\beta $ and $\gamma $ are the Majorana CP violating phases.

The neutrinoless double beta decay restricts the averaged neutrino mass defined by
\be
\la m_\nu\ra_{ee} =\left| \sum_{j=1}^{3}U_{ej}^2m_j\right|
\label{mbetabeta}
\ee
and its experimental upper limits in on-going and near future experiments are listed
in literature \cite{Bernhard} and their sensitivity is in the range
\be
40 < \la m_\nu\ra_{ee} <100 ~~\mbox{meV}.
\ee
This averaged neutrino mass is generalized to
\be
\la m_\nu\ra_{ab} =\left| \sum_{j=1}^{3}U_{aj}U_{bj}m_j\right|,
\label{averagedmass}
\ee
where $a$ and $b$ take $e,~\mu,~\tau$, characterizing other $\Delta L=2$ processes.

On the other hand, the cosmological constraint on neutrino mass has come to
\be
\sum m_i=m_1+m_2+m_3 < 0. 23~~\mbox{eV (Planck+WP+highL+BAO)} ~\cite{Planck}.
\label{CC}
\ee

In the previous papers \cite{F-N1} \cite{F-N2} \cite{F-N3} \cite{F-N4} \cite{F-N5}, we examined the unknown MNS parameters by using experimental data of the time. 
A lot of time passed since then, and these experimental data have been greatly improved and newly measured. 
That is, the present observed values of the lepton mixing angle parameters $s^2_{ij}=\sin^2\theta_{ij}$ are
\bea
s^2_{12}&=&0.32\pm 0.05, \label{theta12}\\
s^2_{23}&=&0.36-0.68\mbox{  for the NH case }, \label{theta23NH}\\
s^2_{23}&=&0.37-0.67\mbox{  for the IH case }, \label{theta23IH}\\
s^2_{13}&=&0.0246_{-0.0076}^{+0.0084}\mbox{  for the NH case }, \label{theta13NH}\\
s^2_{13}&=&0.025\pm 0.008\mbox{  for the IH case }, \label{theta13IH}
\eea
and neutrino mass squared differences $\Delta m_{ij}^2\equiv m_i^2-m_j^2$ are
\bea
\Delta m_{21}^2 &=& 7.62_{-0.50}^{+0.58}\times10^{-5}\mbox{ eV}^2, \label{21} \\
  \Delta m_{31}^2 &=& 2.55_{-0.24}^{+0.19}\times10^{-3}\mbox{eV}^2\mbox{  for the NH case }, \label{31NH}\\
  \Delta m_{13}^2 &=& 2.43_{-0.22}^{+0.21}\times10^{-3}\mbox{eV}^2\mbox{  for the IH case}, \label{31IH}
\eea
at $3\sigma$ level~\cite{Forero} (See also \cite{Fogli}). 
Here the NH(IH) indicates normal (inverted) neutrino mass hierarchy.
In the present paper, by using these new developments we reexamine the constraints of the averaged neutrino masses from the MNS parameters together with the cosmological constraint \bref{CC}.

For the NH case in which $m_1<m_2<m_3$, the neutrino masses are written in terms of $\overline{m_\nu}$  under the unitarity condition $\sum_i |U_{ei}|^2=1$ as \cite{F-N4}

\bea
m_1 & =&
  \sqrt{\overline{m_{\nu}}^2-(1-|U_{e1}|^2) \Delta m_{21}^2-|U_{e3}|^2\Delta m_{32}^2}, \label{eq1220-11},\\
m_2 & =&\sqrt{m_1^2+\Delta m_{21}^2},\\
m_3 & =&\sqrt{m_1^2+\Delta m_{31}^2}\label{eq1220-13}.
\eea
For the IH case in which $m_3<m_1<m_2$, they take the form \cite{F-N4},
\bea
m_3 & =&
  \sqrt{\overline{m_{\nu}}^2-(1-|U_{e3}|^2) \Delta m_{23}^2+|U_{e1}|^2\Delta m_{21}^2}, \label{eq1220-14}\\
m_1 & =&\sqrt{m_3^2+\Delta m_{13}^2}, \label{eq1220-15} \\
m_2 & =&\sqrt{m_3^2+\Delta m_{13}^2+\Delta m_{21}^2}. \label{eq1220-16} 
\eea
In the following numerical analysis, we take the center values of $s^2_{12}$, $s^2_{13}$, $\Delta m_{31}^2$, $\Delta m_{21}^2$ given by \bref{theta12}, \bref{theta13NH}-\bref{31IH}, 
while we vary the values from 0 to $2\pi$ for Majorana CP violating phases $\beta $ and $\gamma $ and for the Dirac CP violating phase $\delta$ too.
As for $s^2_{23}$, we take the lower, center, and upper values shown in \bref{theta23NH} and \bref{theta23IH} for typical values for $s^2_{23}$.

First, following our ealiar work \cite{F-N4}, we derive the neumerical lower limit of $\overline{m_\nu}$. 
In the NH case, the following lower limit of $\overline{m_\nu}$ is derived from \bref{eq1220-11} with $m_1^2\ge0$,
\be
  0.0093~\mbox{eV}=\sqrt{|U_{e2}|^2 \Delta m_{21}^2+|U_{e3}|^2
  \Delta m_{31}^2} <\overline{m_\nu}, 
\label{mbeta_lowerlimit_NH}
\ee
while in the IH case, from \bref{eq1220-14} with $m_3^2\ge0$, we obtain,
\be
  0.049~\mbox{eV}=\sqrt{|U_{e2}|^2 \Delta m_{21}^2+(1-|U_{e3}|^2)
  \Delta m_{13}^2} <\overline{m_\nu}.
\label{mbeta_lowerlimit_IH}
\ee

Next we show the relation between $\overline{m_\nu}$  and $\sum m_i \equiv m_1+m_2+m_3$ in Fig.~1, 
which is obtained by using \bref{eq1220-11}- \bref{eq1220-16}.  
In Fig.~1 we combine this relation with the constraints given by \bref{CC}, \bref{mbeta_lowerlimit_NH} and \bref{mbeta_lowerlimit_IH}, 
so that the upper and lower limits for $\overline{m_\nu}$ and $\sum m_i$ can be derived. 
Namely we obtain
\be
 0.0093 <\overline{m_\nu}<0.072~~\mbox{eV},
\label{mbetalimit NH}
\ee
\be
0.060 <\sum m_i<0.23~~\mbox{eV}.
\label{msumlimit NH}
\ee
in the NH case, and 
\be
0.049 <\overline{m_\nu}< 0.082~~\mbox{eV},
\label{mbetalimit IH}
\ee
\be
0.10 <\sum m_i<0.23~~\mbox{eV}.
\label{msumlimit IH}
\ee
in the IH case.

\begin{center}{\scalebox{0.6}{\includegraphics{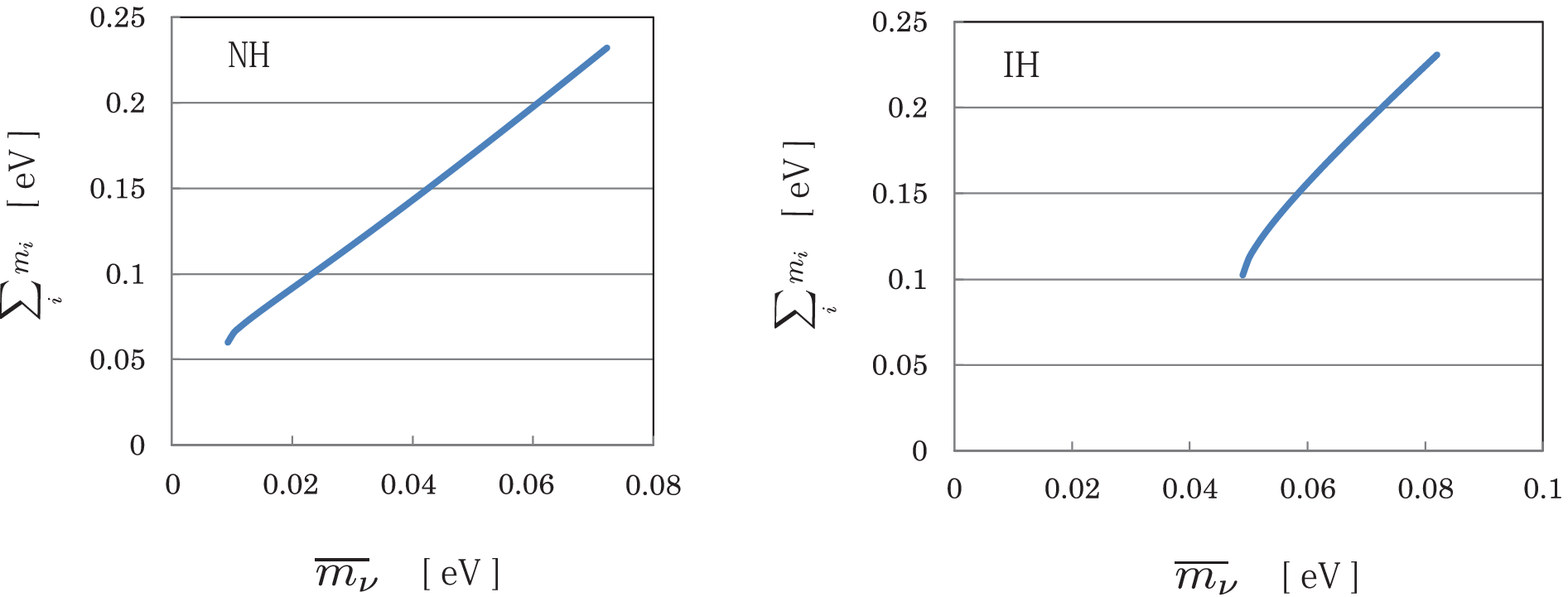}} }\end{center}
\begin{quotation}
{\bf Fig.~1} The relation between $\sum m_i$ and  $\overline{m_\nu}$ in the NH and IH cases.  
The solid curves indicate the relation between $\sum m_i$ and  $\overline{m_\nu}$ in the NH and IH cases. 
We have experimental upper limits given by (8) and theoretical lower limits given by (23) and (24).
\end{quotation}

By substituting \bref{eq1220-11}-\bref{eq1220-13} into \bref{averagedmass} and using \bref{mbetalimit NH} 
for the NH case (\bref{eq1220-14}-\bref{eq1220-16} into \bref{averagedmass} and using \bref{mbetalimit IH} for the IH case), 
we obtain the allowed region in the $\la m_\nu \ra_{ab}$ -$\overline{m_\nu}$ parameter plane 
as shown in the figures from Fig.~2 to Fig.~7. 
These allowed regions are presented as the inside regions bounded by curves, 
which are obtained by varying the values from 0 to $2\pi$ for Majorana CP violating phases $\beta $ and $\gamma $ 
and for the Dirac CP violating phase $\delta$ too. 
Here we take the lower, center, and upper values shown in \bref{theta23NH} and \bref{theta23IH} 
for typical values for $s^2_{23}$, indicating dotted, solid, and dashed curves, respectively in the figures from Fig.~3 to Fig.~7. 
Note that the $\la m_\nu\ra_{ee}$ is independent of $s^2_{23}$.

\begin{center}{\scalebox{0.6}{\includegraphics{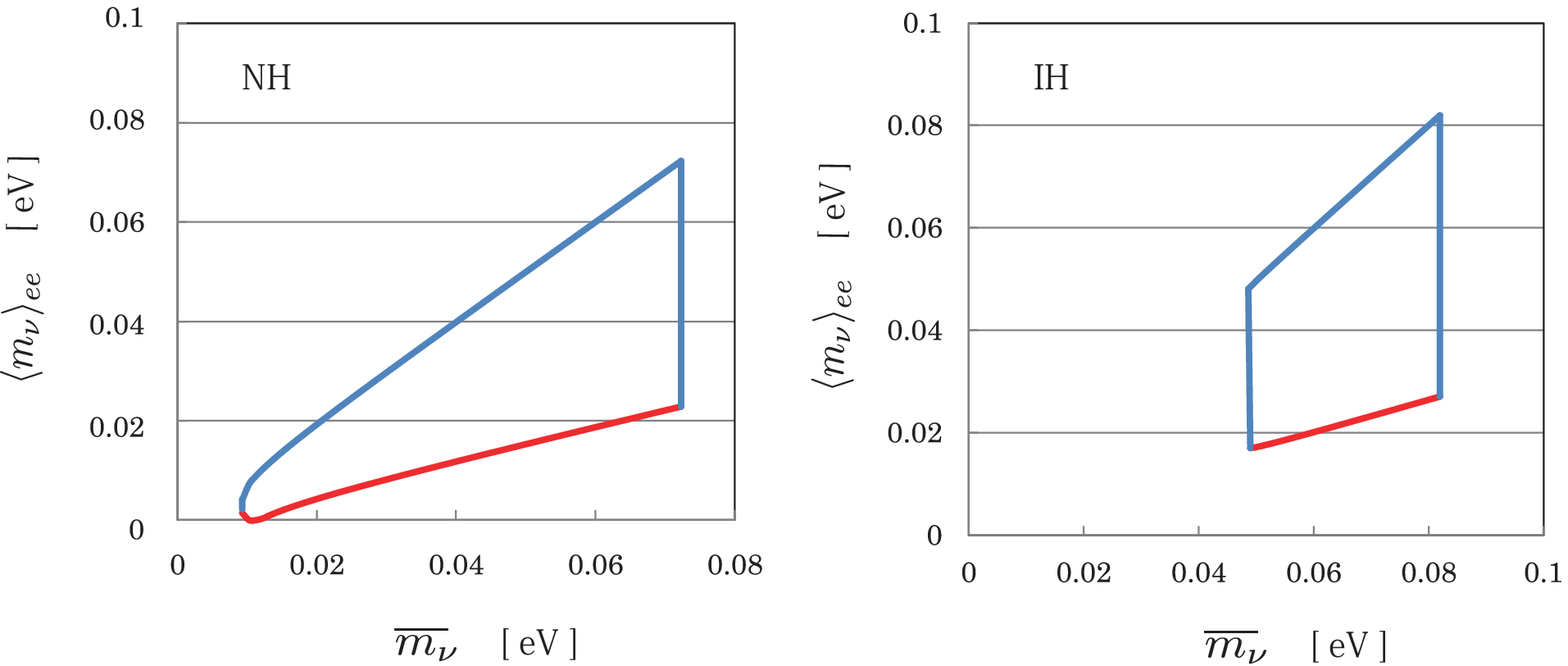}} }\end{center}
\begin{quotation}
{\bf Fig.~2}  The allowed region in the $\langle m_\nu \rangle_{ee}$-$\overline{m_\nu}$ plane in the NH and IH cases. The allowed regions are presented by the inside regions bounded by curves.   
\end{quotation}

\begin{center}{\scalebox{0.6}{\includegraphics{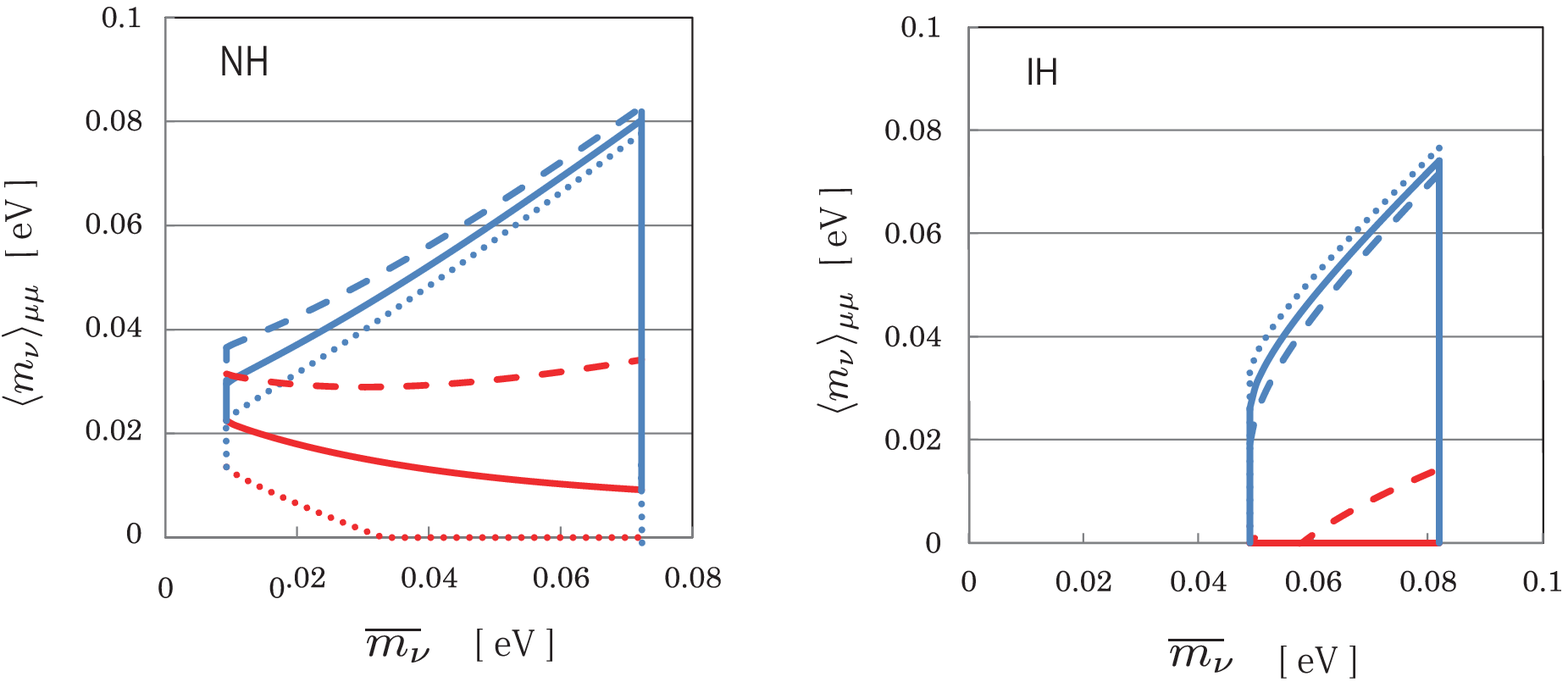}} }\end{center}
\begin{quotation}
{\bf Fig.~3}  The allowed region in the $\langle m_\nu \rangle_{\mu \mu}$-$\overline{m_\nu}$ plane in the NH and IH cases. 
Here and hereafter the dotted, solid, and dashed curves, respectively, correspond to the cases in which 
the lower, center, and upper values shown in \bref{theta23NH} and \bref{theta23IH} are taken for $s^2_{23}$.
\end{quotation}

\begin{center}{\scalebox{0.6}{\includegraphics{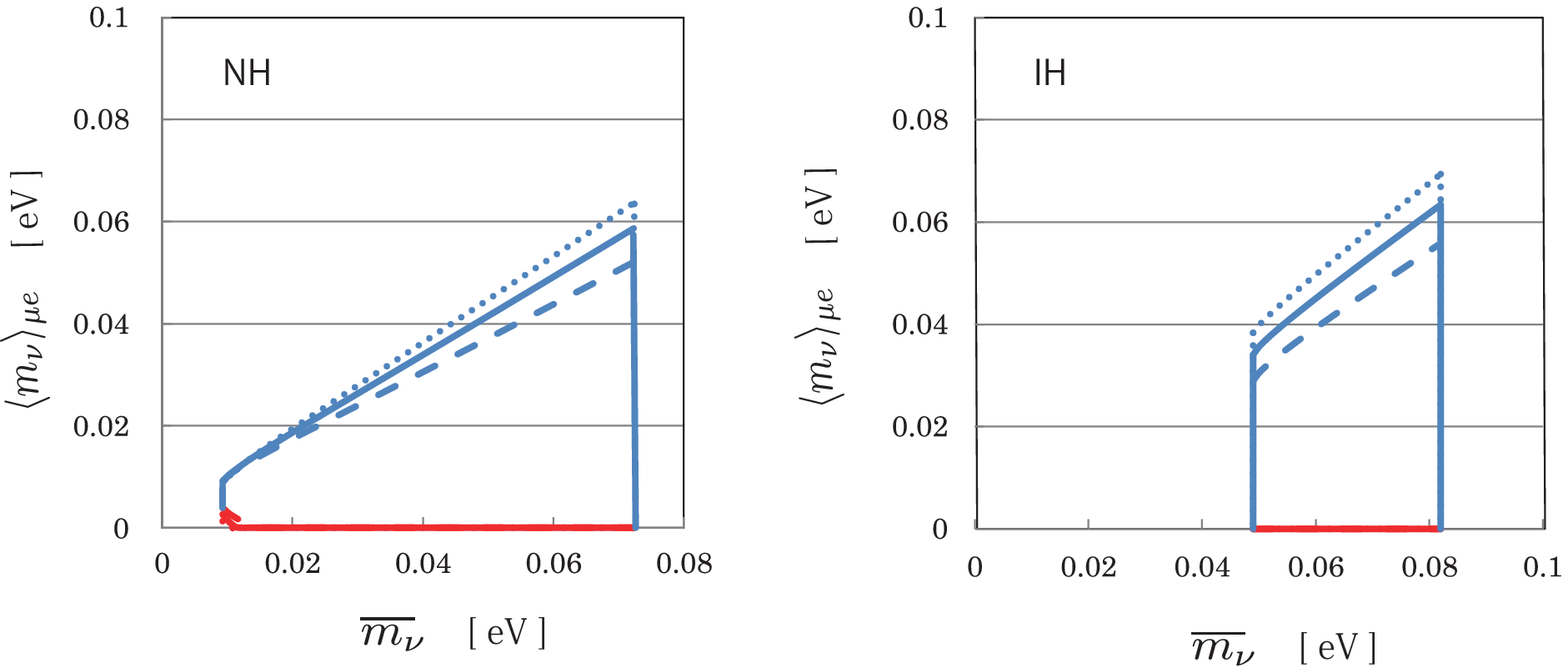}} }\end{center}
\begin{quotation}
{\bf Fig.~4}  The allowed region in the $\langle m_\nu \rangle_{\mu e}$-$\overline{m_\nu}$ plane in the NH and IH cases. 
\end{quotation}

\begin{center}{\scalebox{0.6}{\includegraphics{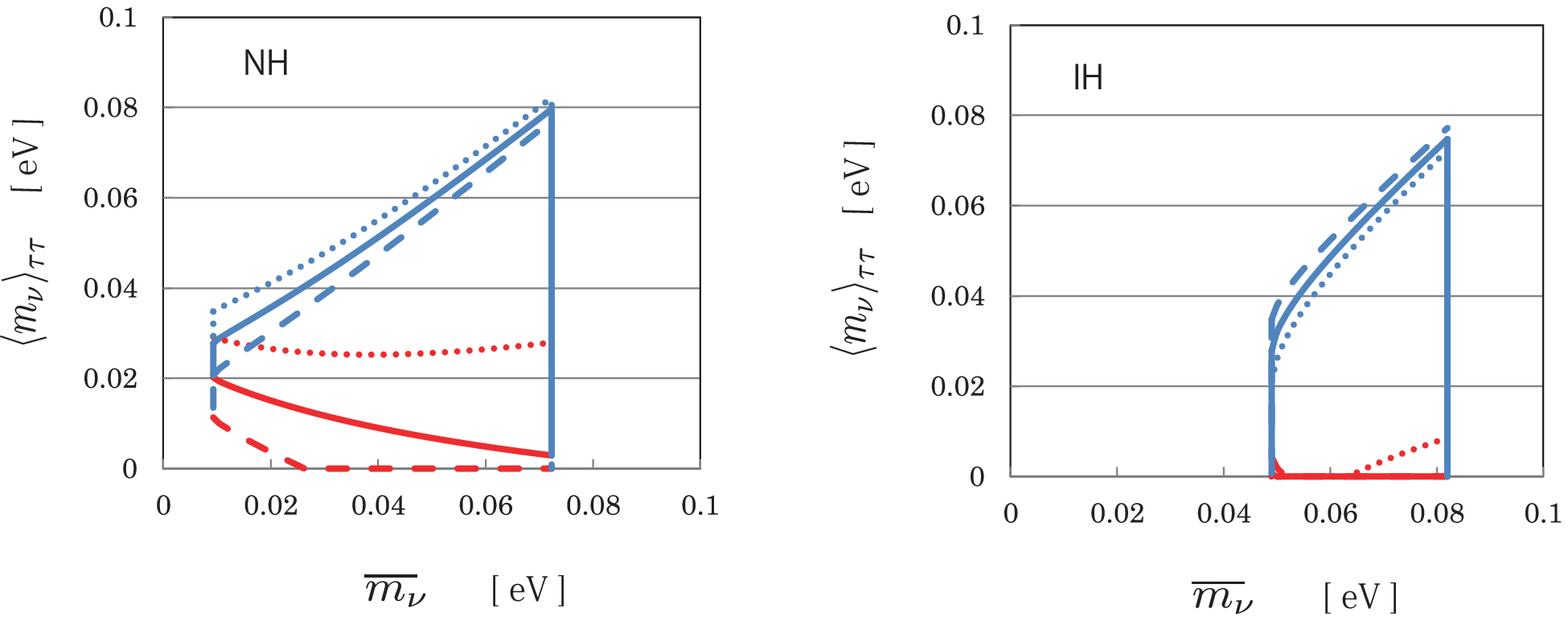}} }\end{center}
\begin{quotation}
{\bf Fig.~5}  The allowed region in the $\langle m_\nu \rangle_{\tau \tau}$-$\overline{m_\nu}$ plane in the NH and IH cases. 
\end{quotation}

\begin{center}{\scalebox{0.6}{\includegraphics{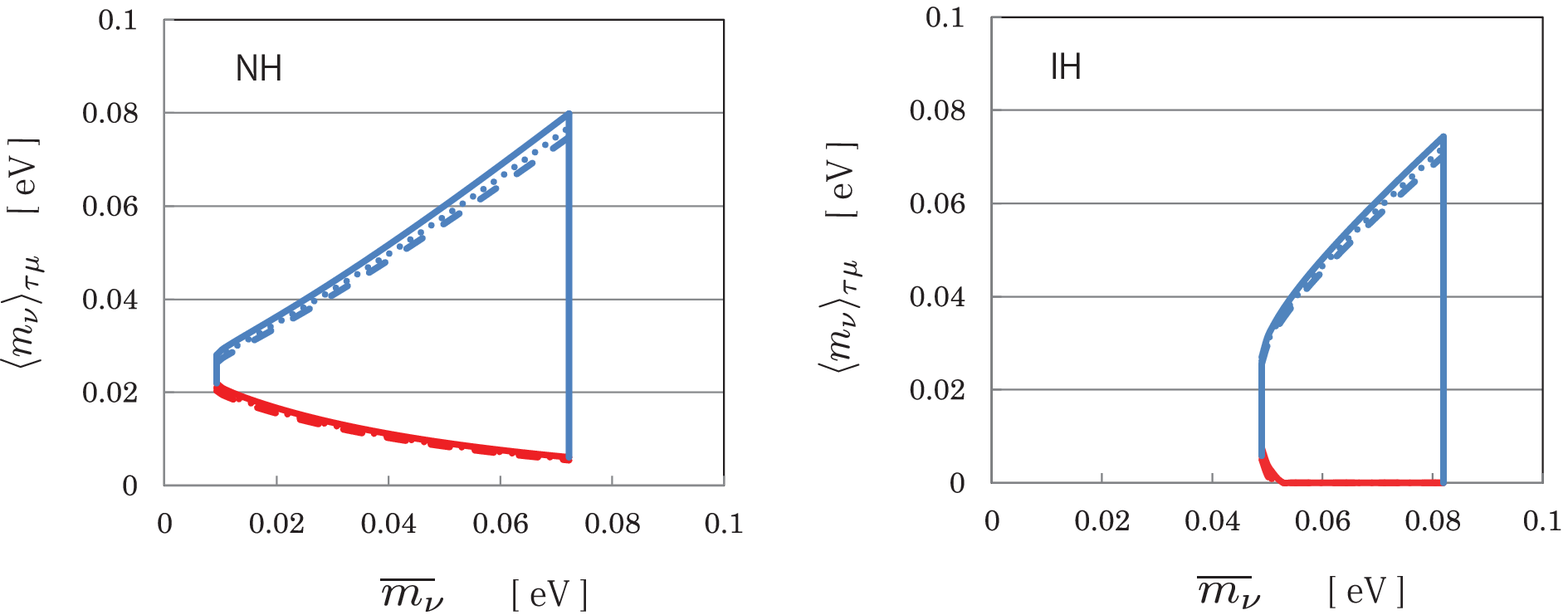}} }\end{center}
\begin{quotation}
{\bf Fig.~6}  The allowed region in the $\langle m_\nu \rangle_{\tau \mu}$-$\overline{m_\nu}$ plane in the NH and IH cases. 
\end{quotation}

\begin{center}{\scalebox{0.6}{\includegraphics{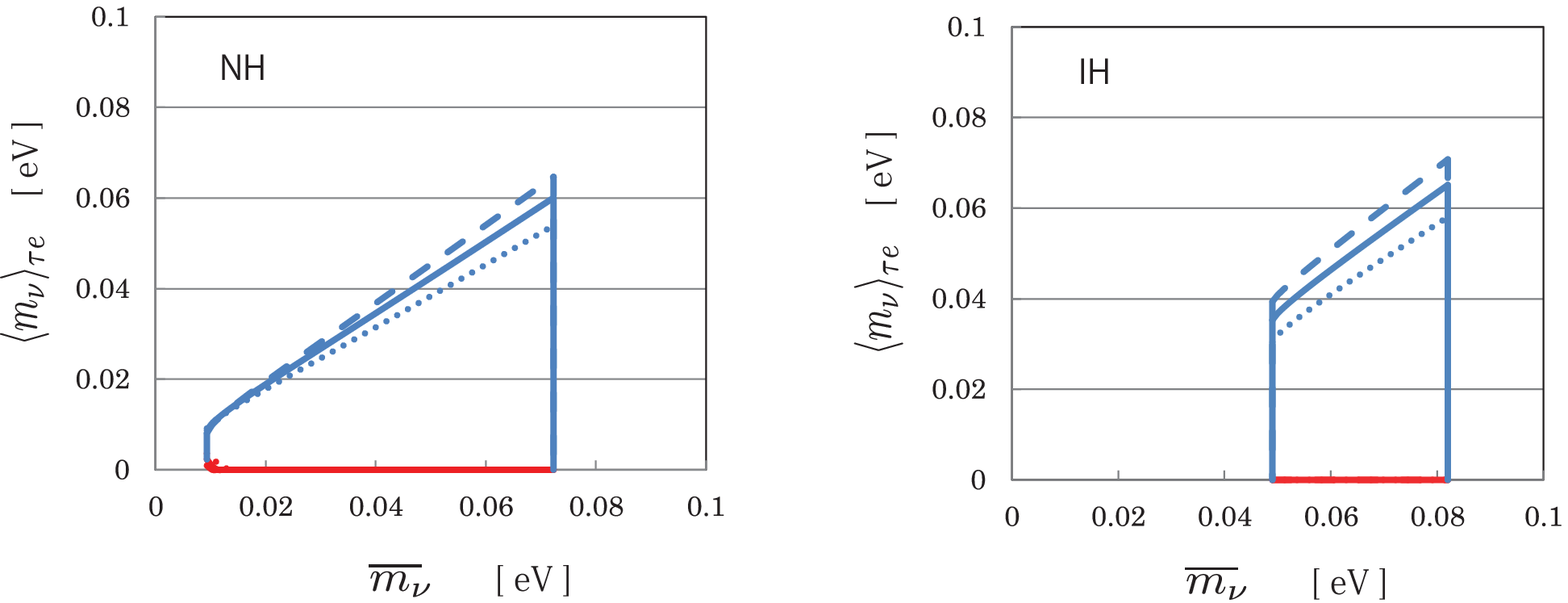}} }\end{center}
\begin{quotation}
{\bf Fig.~7}  The allowed region in the $\langle m_\nu \rangle_{\tau e}$-$\overline{m_\nu}$ plane in the NH and IH cases. 
\end{quotation}

It turns out from  Fig.~2 to Fig.~7 that we have, irrespectively of values of the CP violating phases, 
\bea
&&17~\mbox{meV } \leq \langle m_\nu \rangle_{ee}\leq 80~\mbox{meV for IH case},\\
&&~~5~\mbox{meV }\leq \langle m_\nu \rangle_{\tau \mu}\leq 80~\mbox{meV for NH case},
\label{bound}
\eea
\bea
\langle m_\nu \rangle_{\mu e}\leq 70~\mbox{meV }, \langle m_\nu \rangle_{\mu \mu}\leq 80~\mbox{meV }, \\
\langle m_\nu \rangle_{\tau e}\leq 80~\mbox{meV }, \langle m_\nu \rangle_{\tau \mu}\leq 80~\mbox{meV }.
\label{bound2}
\eea
These results indicate that $\langle m_\nu \rangle_{ee}$ in the IH case and $\langle m_\nu \rangle_{\tau \mu}$ in the NH case 
have the lower bounds, and that all the averaged masses $\la m_\nu\ra _{ab}$ take values at most $80~\mbox{meV}$. 

Let us consider the implications of our findings.
The lower bound of $\langle m_\nu \rangle_{ee}$ is rather near to the sensitivity of near future experiments. 
So if $\langle m_\nu \rangle_{ee}$ will not be found above this bound, the IH scenario is disproved. 
The experimental constraints of the general averaged masses $\la m_\nu\ra _{ab}$ other than $\la m_\nu\ra _{ee}$ are not stringent. 
For lepton number violating $\Delta L=2$ process, the contribution from the standard model + massive Majorana neutrinos seems to be much smaller than the present experimental upper bound. However we stress that  our prediction $\langle m_\nu \rangle_{\tau \mu}\geq 5~\mbox{meV for NH}$ leads to a lower limit for the amplitude for some $\Delta L=2$ process which proportional to $\langle m_\nu \rangle_{\tau \mu}$.

The upper bound of $\la m_\nu\ra_{\mu\mu}<80~\mbox{meV}$ leads to theoretical branching ratio \cite{Abad}
\be
\frac{\Gamma(K^-\rightarrow \pi^+\mu^-\mu^-)}{K^-\rightarrow all}<5\times 10^{-35}
\ee
which is far from the experimental bound~ \cite{PDG} given by
\be
\frac{\Gamma(K^-\rightarrow \pi^+\mu^-\mu^-)}{K^-\rightarrow all}<1.1 \times 10^{-9} ~(90\% C.L.)).
\label{Kbound}
\ee
However, this estimation is essentially based on the Standard Model + the massive Majorana neutrino.
For the $\Delta L=2$ process of a triplet Higgs model, for example, the triplet Higgs takes the important role 
whose interaction is written as
\be
{\mathcal L}_{\Delta}^{} =
-\sqrt2(h_M^\dag U)_{\ell i}\overline{\ell_L^{}}N_i^c\Delta^- 
-(h_M^\dag)_{\ell\ell'}\overline{\ell_L^{}}{{\ell'}_L}^c\Delta^{--}
+\text{H.c.}
\ee
Here
\be
|h_{M,ab}|=|Um_\nu^\text{diag}U^T)_{ab}|/(\sqrt2v_\Delta)\equiv \la m_\nu\ra_{ab}/(\sqrt2v_\Delta^{}), 
\ee
and 
$N_i(i=1,2,3)$ represent Majorana neutrinos which satisfy the conditions $N_i=N_i^c=C\overline{N_i}^T$. 
Also, upper bound of the $BR(\tau^-\rightarrow \mu^+\pi^-\pi^-)$ is predicted \cite{Quintero} as
\be
BR(\tau^-\rightarrow \mu^+\pi^-\pi^-)<4.41\times 10^{-23}, 
\ee
whereas the experimental upper limit is $3.9 \times 10^{-8}$ \cite{Miyazaki}. 
The difference between the theoretical and experimental bounds is also still large.
However, SUSY with R-parity violation may give much larger contribution \cite{Martin}. 

In conclusion we have discussed the upper and lower bounds on the averaged neutrino masses $\la m_\nu\ra _{ab}$ 
in the light of the latest data of the neutrino oscillation and the cosmological constraint on neutrino mass. 
Note that the efective neutrino mass $\langle m_\nu \rangle_{ee}$ has been updated in the literature \cite{Rodejohann} too, 
and recently the upper bounds on $\la m_\nu\ra _{ab}$ are also studied by Xing and Zhou \cite{Xing-Zhou}.

\section*{Acknowledgements}
We are grateful to M. Hasegawa (KEK) for useful comments.
The work of T.F.\ is supported in part by the Grant-in-Aid for Science Research
from the Ministry of Education, Science and Culture of Japan
(No.~21104004).

\end{document}